\documentclass{aa}
\usepackage{graphicx}
\usepackage{natbib}
\usepackage{txfonts}
\begin{document}
\newcommand{\hess}{H.E.S.S.}
\newcommand{\vhe}{V\textsc{HE}}
\newcommand{\vela}{\object{RX\,J0852.0$-$4622}}
\newcommand{\ax}{\object{AX\,J0851.9$-$4617.4}}
\newcommand{\sax}{\object{SAX\,J0852.0$-$4615}}
\newcommand{\cxou}{\object{CXOU\,J085201.4$-$461753}}
\newcommand{\gammaS}{$\gamma$}
\newcommand{\PhotonIndex}{$\Gamma = 2.1 \pm 0.1_{\mathrm{stat}}$}
\newcommand{\Flux}{$\Phi(E>1\,\mbox{TeV}) = (1.9 \pm 0.3_{\mathrm{stat}}) \times 10^{-11} \mbox{cm}^{-2}\mbox{s}^{-1}$}
\newcommand{\Norm}{$\varphi_{1\,\mathrm{TeV}} = (2.1 \pm 0.2_{\mathrm{stat}}) \times 10^{-11} \mbox{cm}^{-2}\mbox{s}^{-1}\mbox{TeV}^{-1}$}
\newcommand{\NormSys}{$\varphi_{1\,\mathrm{TeV}} = (2.1 \pm
0.2_{\mathrm{stat}} \pm 0.6_{\mathrm{syst}}) \times 10^{-11} \mbox{cm}^{-2}\mbox{s}^{-1}\mbox{TeV}^{-1}$}
\newcommand{\PhotonIndexSys}{$\Gamma = 2.1 \pm 0.1_{\mathrm{stat}} \pm 0.2_{\mathrm{syst}}$}
\newcommand{\FluxSys}{$\Phi(E>1\,\mbox{TeV}) = (1.9 \pm
0.3_{\mathrm{stat}} \pm 0.6_{\mathrm{syst}}) \times 10^{-11}
\mbox{cm}^{-2}\mbox{s}^{-1}$}
\newcommand{\AXUL}{$1.3 \times 10^{-12}\,\mathrm{cm}^{-2} \mathrm{s}^{-1}$}
\newcommand{\excess}{($700 \pm 60$)}
\newcommand{\Non}{2406}
\newcommand{\Noff}{2541}
\newcommand{\normAlpha}{0.671}
\newcommand{\sig}{12\,$\sigma$}
\newcommand{\rate}{$(3.7 \pm 0.3)$\,min$^{-1}$}
\newcommand{\Eth}{400\,GeV}
\newcommand{\wGamma}{$w_\gamma(1-10\,\mathrm{TeV}) =
\int_{1\,\rm{TeV}}^{10\,\rm{TeV}} E \varphi(E) \mathrm{d}E \approx 7 \times 10^{-11} \mathrm{erg}\, \mathrm{cm}^{-2} \mathrm{s}^{-1}$}
\newcommand{\LGamma}{$L_\gamma (1-10\,\mathrm{TeV})=4 \pi d^2 w_\gamma(1-10\,\mathrm{TeV}) \approx 3 \times 10^{32}(d/200\,\mathrm{pc})^2 \,\mathrm{erg/s}$}
\newcommand{\Wprot}{$W(10-100\,\mathrm{TeV}) \approx t_{pp \rightarrow
\pi^0} \times L_\gamma(1-10\,\mathrm{TeV}) \approx 1.5 \times 10^{48} (d/200\,\mathrm{pc})^2 (n/1\,\mathrm{cm}^{-3})^{-1}\,\mathrm{erg}$}
\newcommand{\Wtot}{$W_{\mathrm{tot}} \approx 10^{49} (d/200\,\mathrm{pc})^2 (n/1\,\mathrm{cm}^{-3})^{-1} \mathrm{erg}$}

\title{Detection of TeV \gammaS -ray Emission from the Shell-Type Supernova Remnant \vela\ with H.E.S.S.}

\author{F. Aharonian\inst{1}
 \and A.G.~Akhperjanian \inst{2}
 \and A.R.~Bazer-Bachi \inst{3}
 \and M.~Beilicke \inst{4}
 \and W.~Benbow \inst{1}
 \and D.~Berge \inst{1}
 \and K.~Bernl\"ohr \inst{1,5}
 \and C.~Boisson \inst{6}
 \and O.~Bolz \inst{1}
 \and V.~Borrel \inst{3}
 \and I.~Braun \inst{1}
 \and F.~Breitling \inst{5}
 \and A.M.~Brown \inst{7}
 \and P.M.~Chadwick \inst{7}
 \and L.-M.~Chounet \inst{8}
 \and R.~Cornils \inst{4}
 \and L.~Costamante \inst{1,20}
 \and B.~Degrange \inst{8}
 \and H.J.~Dickinson \inst{7}
 \and A.~Djannati-Ata\"{\i} \inst{9}
 \and L.O'C.~Drury \inst{10}
 \and G.~Dubus \inst{8}
 \and D.~Emmanoulopoulos \inst{11}
 \and P.~Espigat \inst{9}
 \and F.~Feinstein \inst{12}
 \and G.~Fontaine \inst{8}
 \and Y.~Fuchs \inst{13}
 \and S.~Funk \inst{1}
 \and Y.A.~Gallant \inst{12}
 \and B.~Giebels \inst{8}
 \and S.~Gillessen \inst{1}
 \and J.F.~Glicenstein \inst{14}
 \and P.~Goret \inst{14}
 \and C.~Hadjichristidis \inst{7}
 \and M.~Hauser \inst{11}
 \and G.~Heinzelmann \inst{4}
 \and G.~Henri \inst{13}
 \and G.~Hermann \inst{1}
 \and J.A.~Hinton \inst{1}
 \and W.~Hofmann \inst{1}
 \and M.~Holleran \inst{15}
 \and D.~Horns \inst{1}
 \and A.~Jacholkowska \inst{12}
 \and O.C.~de~Jager \inst{15}
 \and B.~Kh\'elifi \inst{1}
 \and Nu.~Komin \inst{5}
 \and A.~Konopelko \inst{1,5}
 \and I.J.~Latham \inst{7}
 \and R.~Le Gallou \inst{7}
 \and A.~Lemi\`ere \inst{9}
 \and M.~Lemoine-Goumard \inst{8}
 \and N.~Leroy \inst{8}
 \and T.~Lohse \inst{5}
 \and J.M.~Martin \inst{6}
 \and O.~Martineau-Huynh \inst{16}
 \and A.~Marcowith \inst{3}
 \and C.~Masterson \inst{1,20}
 \and T.J.L.~McComb \inst{7}
 \and M.~de~Naurois \inst{16}
 \and S.J.~Nolan \inst{7}
 \and A.~Noutsos \inst{7}
 \and K.J.~Orford \inst{7}
 \and J.L.~Osborne \inst{7}
 \and M.~Ouchrif \inst{16,20}
 \and M.~Panter \inst{1}
 \and G.~Pelletier \inst{13}
 \and S.~Pita \inst{9}
 \and G.~P\"uhlhofer \inst{1,11}
 \and M.~Punch \inst{9}
 \and B.C.~Raubenheimer \inst{15}
 \and M.~Raue \inst{4}
 \and J.~Raux \inst{16}
 \and S.M.~Rayner \inst{7}
 \and A.~Reimer \inst{17}
 \and O.~Reimer \inst{17}
 \and J.~Ripken \inst{4}
 \and L.~Rob \inst{18}
 \and L.~Rolland \inst{16}
 \and G.~Rowell \inst{1}
 \and V.~Sahakian \inst{2}
 \and L.~Saug\'e \inst{13}
 \and S.~Schlenker \inst{5}
 \and R.~Schlickeiser \inst{17}
 \and C.~Schuster \inst{17}
 \and U.~Schwanke \inst{5}
 \and M.~Siewert \inst{17}
 \and H.~Sol \inst{6}
 \and D.~Spangler \inst{7}
 \and R.~Steenkamp \inst{19}
 \and C.~Stegmann \inst{5}
 \and J.-P.~Tavernet \inst{16}
 \and R.~Terrier \inst{9}
 \and C.G.~Th\'eoret \inst{9}
 \and M.~Tluczykont \inst{8,20}
 \and G.~Vasileiadis \inst{12}
 \and C.~Venter \inst{15}
 \and P.~Vincent \inst{16}
 \and H.J.~V\"olk \inst{1}
 \and S.J.~Wagner \inst{11}}

\institute{
Max-Planck-Institut f\"ur Kernphysik, Heidelberg, Germany
\and
 Yerevan Physics Institute, Armenia
\and
Centre d'Etude Spatiale des Rayonnements, CNRS/UPS, Toulouse, France
\and
Universit\"at Hamburg, Institut f\"ur Experimentalphysik, Germany
\and
Institut f\"ur Physik, Humboldt-Universit\"at zu Berlin, Germany
\and
LUTH, UMR 8102 du CNRS, Observatoire de Paris, Section de Meudon, France
\and
University of Durham, Department of Physics, U.K.
\and
Laboratoire Leprince-Ringuet, IN2P3/CNRS,
Ecole Polytechnique, Palaiseau, France
\and
APC, Paris, France 
\thanks{UMR 7164 (CNRS, Universit\'e Paris VII, CEA, Observatoire de Paris)}
\and
Dublin Institute for Advanced Studies, Ireland
\and
Landessternwarte, K\"onigstuhl, Heidelberg, Germany
\and
Laboratoire de Physique Th\'eorique et Astroparticules, IN2P3/CNRS,
Universit\'e Montpellier II, France
\and
Laboratoire d'Astrophysique de Grenoble, INSU/CNRS, Universit\'e Joseph Fourier, France 
\and
DAPNIA/DSM/CEA, CE Saclay,
Gif-sur-Yvette, France
\and
Unit for Space Physics, North-West University, Potchefstroom,
    South Africa
\and
Laboratoire de Physique Nucl\'eaire et de Hautes Energies, IN2P3/CNRS, Universit\'es
Paris VI \& VII, France
\and
Institut f\"ur Theoretische Physik, Lehrstuhl IV,
    Ruhr-Universit\"at Bochum, Germany
\and
Institute of Particle and Nuclear Physics, Charles University, Prague, Czech Republic
\and
University of Namibia, Windhoek, Namibia
\and
European Associated Laboratory for Gamma-Ray Astronomy, jointly
supported by CNRS and MPG}

\offprints{Nukri Komin, \email{komin@physik.hu-berlin.de}}

\date{Received ? / Accepted ?}

\abstract{
We report the detection of TeV \gammaS -rays from the shell-type
supernova remnant \vela\ with data of 3.2\,h of live time recorded with
\hess\ in February 2004. An excess of \excess\ events from the whole remnant
with a significance of \sig\ was found. The observed emission region
is clearly extended with a radius of the order of $1\degr$ and the
spatial distribution of the signal correlates with X-ray
observations. The spectrum in the energy range between 500\,GeV and
15\,TeV is well described by a power law with a photon index of
\PhotonIndexSys\ and a differential flux at 1\,TeV of
\NormSys . The integral flux above 1\,TeV was
measured to be \FluxSys, which is at the level of the flux of the Crab
nebula at these energies. More data are needed to draw firm
conclusions on the magnetic field in the remnant and the type of the
particle population creating the TeV \gammaS -rays.

   \keywords{Gamma rays: observations -- supernovae: individual: 
   RX\,J0852.0$-$4622 (G266.2$-$1.2)} }

\authorrunning{F.\ Aharonian et al. (H.E.S.S. collaboration)}
\titlerunning{Detection of TeV emission from \vela\ with H.E.S.S.}

   \maketitle
%

\section{Introduction}

\vela\ (also called \object{G266.2$-$1.2}) is a
young shell-type supernova remnant (SNR) in the line of sight to the
Vela SNR.  The observed X-ray emission of \vela\ extends over a
roughly circular region with a diameter of $\sim \! 2\degr$ with a
brightening towards the north-western, western and southern part of
the shell and towards the centre.  The observed X-ray spectrum is
clearly dominated by a continuum which indicates a non-thermal origin
of the emission \citep{Aschenbach,Tsunemi,VelaJrASCA,XMM}. Deep
X-ray observations with the ASCA, CHANDRA and BeppoSAX satellites
revealed a compact source in the central region of \vela :
\ax\ (\cxou )
\citep{VelaJrASCA,VelaJrBeppoSAX,VelaJr2,CHANDRACentral}. This source has been
suggested to be a neutron star, supported by
the detection of an coincident H$\alpha$ nebula \citep{HaCenter}.
An association of the neutron star candidate with \vela\ would point
to a core-collapse supernova. However, recent X-ray
observations suggest that \vela\ is the result of a sub-Chandrasekhar
type Ia supernova explosion \citep{XMM}, which would imply that the
compact object is not related to \vela.  Radio observations show
only weak emission from the shell and no emission from the centre
\citep{VelaJrRadio}.  The age and distance of \vela\ were calculated
from the diameter seen in X-rays and the flux of $^{44}$Ti lines to be
$680\pm100$\,years and $\sim \! 200$\,pc with upper limits of
1100\,years and 500\,pc, respectively \citep{Progenitor,Iyudin}. An
age between 630 and 970\,years was estimated by \citet{Tsunemi} based
on the observation of Ca lines. These estimates for distance and age
would imply that \vela\ is one of the closest supernovae in recent
history, whereas \citet{VelaJrASCA} argue that \vela\ might be located
near the Vela Molecular Ridge at a much larger distance of 1--2\,kpc.

Shell-type SNRs with non-thermal X-ray emission are prime candidates
for accelerating cosmic rays up to very high energies
\citep{SN1006_ASCA,RXJ1713_ASCA,Voelk}. Their detection in \vhe\
\gammaS -rays is expected to be possible with modern atmospheric
Cherenkov telescopes \citep{DruryAharonianVoelk94}, and to provide
insight into the underlying acceleration mechanisms. So far, only one
of these SNRs, \object{RX\,J1713.7$-$3946}, was detected by two
independent experiments \citep{CangarooRXJ,Enomoto,Nature} employing
the imaging atmospheric Cherenkov technique. The CANGAROO
collaboration detected \gammaS -ray emission from the north-western
part of the \vela\ SNR \citep{Cangaroo2}. Here we report on the
detection of the entire SNR \vela\ by \hess\ in a short observation
campaign.

\hess\ (High Energy Stereoscopic System) is an array of
four imaging Cherenkov telescopes dedicated to the detection of
\vhe\ \gammaS -rays with energies above 100\,GeV \citep{Performance}. Each
telescope has a tessellated mirror with an area of 107\,m$^2$
\citep{StatusOptics,StatusOptics2} and a camera consisting of 960
photomultiplier tubes \citep{StatusCamera}.  The \hess\ array can
detect point sources at flux levels of about 1\% of the Crab nebula flux
with a significance of 5\,$\sigma$ in 25\,h of observation
\citep{Performance}. \hess\ is currently the most sensitive instrument
to observe \vhe\ \gammaS\ -ray sources. With its angular resolution of
better than $0.1\degr$ per event and its large field of view
($5\degr$) it is additionally in an ideal position to unravel the
\gammaS -ray morphology of extended sources.

\section{Data Set and Analysis Technique}

\vela\ was observed by \hess\ for 4.5\,h in February 2004 in standard
operation mode using all four telescopes and a trigger requiring the
simultaneous observation of an air shower by at least two telescopes
\citep{CentralTrigger}. A run quality selection based on weather
conditions and system monitoring was applied. The selected data,
which were taken at zenith angles between $22\degr$ and $30\degr$ with
a mean of $25\degr$, represent a dead-time-corrected total exposure
time (live time) of 3.16\,h. Due to technical problems during data
taking, the data of only three telescopes could be included in the
analysis.

The background level was estimated from off-source runs, observing sky
regions where no \gammaS -ray sources are known. Off-source data were
recorded between April and June 2004 at zenith angles between
$13\degr$ and $34\degr$ (mean: $25\degr$) with a live time of 4.71\,h
(data set OS1). Another off-source data set (referred to as OS2),
taken at a different sky position with less statistics and a possible
contamination from a \gammaS -ray source, served to verify the results
obtained with OS1. The OS2 data set was recorded between January and
March 2004 at zenith angles in the range of $22\degr$ to $32\degr$
with a mean of $27\degr$. The data of one telescope were excluded from
the analysis of both off-source data sets to match the experimental
setup of the on-source data set.

In order to reject most of the low-energy cosmic rays (CR), only
camera images with intensities of more than 200 photo electrons were
considered for shower reconstruction. For further reduction of the CR
background, cuts on scaled image parameters were applied. These cuts
were optimised using Monte Carlo simulations for point-like sources
with a flux on the level of 10\% of the Crab flux. The directions of
the air showers were reconstructed from shower images in different
cameras and the \gammaS -ray energy was determined from the image
intensity and the shower geometry with a typical relative resolution
of $\sim \! 15\%$. The energy threshold after all cuts is about \Eth
. A detailed description of the analysis technique can be found in
\citet{PKS2155}.

\begin{figure}
\centering
\includegraphics[width = \hsize]{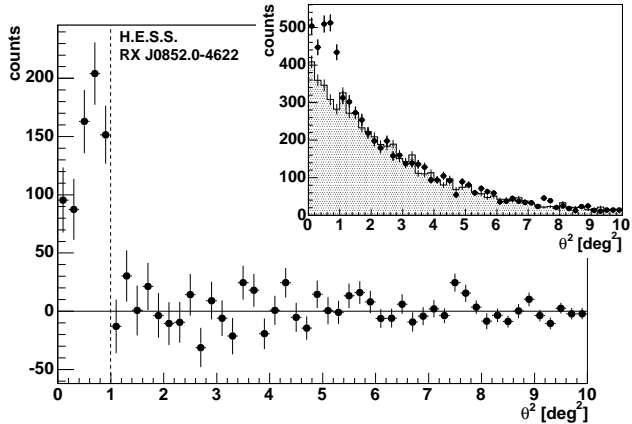}
\caption{
$\theta^2$-plot of the excess of \gammaS -ray like events from
\vela , where $\theta$ is the reconstructed angular distance to
the centre of the SNR. The applied $\theta^2$-cut of 1~deg$^2$ is
denoted by the dashed line. The inset shows the on-source data (data
points) and off-source data (data set OS1, shaded histogram),
normalised using their live-time ratio. The error bars denote
$\pm1\sigma$ statistical errors. }
\label{Fig:ThetaSqr} 
\end{figure}

\begin{figure} 
\centering 
\includegraphics[width = \hsize]{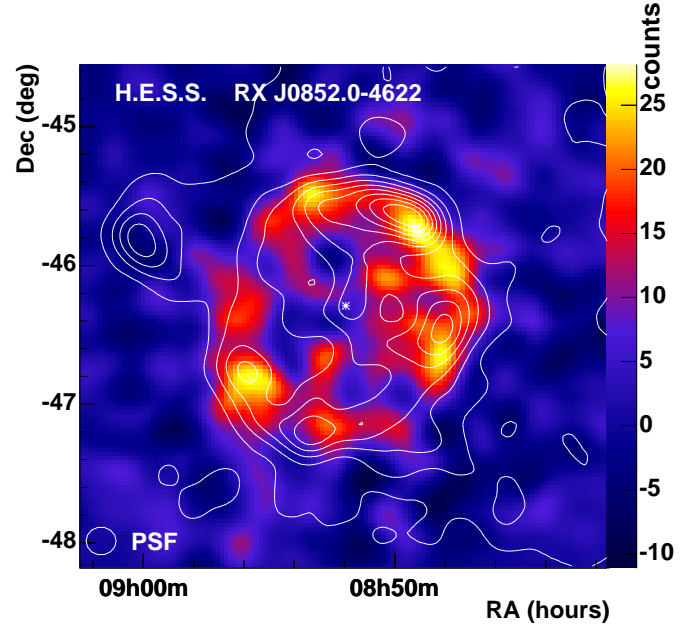} 
\caption{
Count map of \gammaS -rays from the direction of \vela\ after
background subtraction. The data are smoothed with a Gaussian ($\sigma
= 0.1\degr$) representing the angular resolution of the
instrument. The point spread function (PSF) is indicated by a
circle. \gammaS -ray features smaller than the PSF should not be
considered as real. The lines denote equidistant contours of smoothed
($\sigma = 0.1\degr$) X-ray data from the ROSAT All Sky Survey, with
energies restricted to above 1.3\,keV. The position of the neutron
star candidate \ax\ is marked with an asterisk. The axes show J2000.0
equatorial coordinates.}
\label{Fig:Excess}
\end{figure}

\begin{figure}
\centering
\includegraphics[width = \hsize]{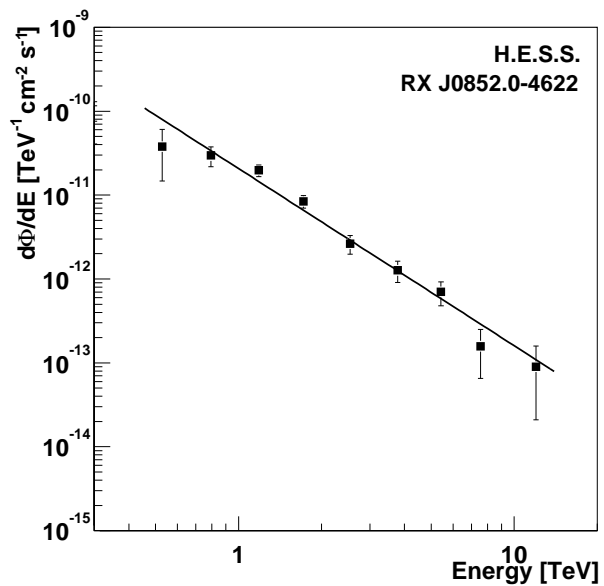}
\caption{
Differential photon flux spectrum of the \gammaS -rays from the
direction of the whole SNR. The solid line is the result of a power
law fit. The error bars denote $\pm1\sigma$ statistical errors.}
\label{Fig:Spectrum}
\end{figure}

\section{Results}

Figure~\ref{Fig:ThetaSqr} shows the radial distribution of the excess
of \gammaS -rays from \vela\ as a function of the reconstructed
squared angular distance, $\theta^2$, from the nominal centre of the
SNR (RA 8h52$\fm$0, Dec $-46\degr 22\arcmin$). The excess was obtained
by subtracting the live-time-normalised background from the on-source
data. The inset of Fig.~\ref{Fig:ThetaSqr} shows good agreement
between on-source and off-source data in the range above
$1~\mathrm{deg}^2$. The distributions are not flat outside the signal
region since the instrument's acceptance drops off towards greater
values of $\theta^2$. In the region $\theta^2
\le 1~\mathrm{deg}^2$, approximately corresponding to the X-ray radius of the SNR, a
clear excess of \excess ~events corresponding to a photon rate of
\rate\ was found. The significance of the signal is \sig, calculated
from \Non\ on events and \Noff\ off events with a normalisation factor
of \normAlpha\ using formula (17) of \citet{LiMa}. The
$\theta^2$-distribution of the excess is much wider than the
distribution measured for point-like sources \citep{PKS2155}. The
source is clearly extended, with a radius of the order of $1\degr$. To
address the question of emission from a compact central object, the
central region of the SNR was tested for the presence of a point-like
source by applying a point-source cut ($\theta^2 \le
0.02~\mathrm{deg}^2$) around \ax . No significant excess was found;
the upper limit (99.9\% confidence level) on the integral photon flux
above 1\,TeV is \AXUL.

A sky map of the excess is displayed in Fig.~\ref{Fig:Excess}. No
correction for the instrument's acceptance, which drops by 23\%
towards the source boundary at $1^\circ$, was applied. An excess of
\gammaS -rays from an extended region is visible. The overlaid contour
plot was derived from X-ray data taken in scanning mode with the PSPC
detector aboard the ROSAT satellite \citep{ROSAT}, restricting the
photon energies to above 1.3\,keV in analogy to the original detection
by \citet{Aschenbach}. We note that the X-ray data are contaminated
with emission from the Vela SNR and \object{RCW\,37} (east of
\vela). Ignoring this and the fact that the \gammaS -ray data were
not corrected for acceptance and exposure, the correlation coefficient
between the
\gammaS -ray counts and X-ray counts in bins of $0.3\degr
\times 0.3\degr$ size was found to be $0.67 \pm 0.05$. The use of ASCA
X-ray data \citep{VelaJrASCA}, which does not cover the complete SNR,
yields a very similar correlation coefficient. More detailed studies
of the \gammaS -ray morphology have to await future high-statistics
observations.

The differential photon flux spectrum of the \gammaS -ray emission from the
whole SNR is shown in Fig.~\ref{Fig:Spectrum}. A power law
\[
\varphi(E) = \frac{\mathrm{d}\Phi}{\mathrm{d}E} = \varphi_{1\,\mathrm{TeV}} \cdot \left(\frac{E}{1\,\mbox{TeV}}\right)^{-\Gamma}
\] 
was fitted to the data points (solid line in Fig.~\ref{Fig:Spectrum})
with a $\chi^2 / \mathrm{d.o.f.} = 10/7$ and results in a photon index
of \PhotonIndex\ and a differential flux at 1\,TeV of \Norm. The
corresponding integral flux above 1\,TeV is \Flux\ which is of the
order of the Crab flux at these energies \citep{HESSCrab}. As a
cross-check, the analysis was repeated using data set OS2 for
background estimation and compatible results were obtained.

Systematic uncertainties were estimated by varying cuts and the
binning in the energy spectrum determination, and by repeating the
analysis with different background data sets (OS1, OS2 and subsamples
of OS1) and with different atmosphere models in the simulation. The
systematic errors are $\sim 0.2$ for the photon index and $\sim 30\%$
for both the differential flux at 1\,TeV and the integral flux.

\section{Discussion}

\vela\ is the second shell-type SNR which has been spatially resolved
at TeV energies, following the H.E.S.S.\ detection of
\object{RX\,J1713.7$-$3946} \citep{Nature}. Similar to
\object{RX\,J1713.7$-$3946}, \vela\ is a weak radio source and was
initially discovered in X-ray observations. There are some other
similarities between these two SNRs, in particular: (i) in both
sources the non-thermal X-ray component strongly dominates over the
thermal component and (ii) both are strong sources of extended TeV
emission spatially correlated with X-rays.

There are two basic mechanisms for TeV \gammaS -ray production in
young SNRs -- inverse Compton scattering (IC) of multi-TeV electrons
on photons of the cosmic microwave background (CMB) and other target photon
fields, and $\pi^0$-decay \gammaS -rays from inelastic interactions of
accelerated protons with ambient gas. The measured \gammaS -ray
flux spectrum of \vela\ translates into an energy flux between 1 and
10\,TeV of \wGamma, which is quite close to the X-ray energy flux of
the entire remnant of $w_{\rm{X}}(0.5-10\,\mathrm{keV}) \sim \! 
10^{-10} \mathrm{erg}\,\mathrm{cm}^{-2} \mathrm{s}^{-1}$
\citep{VelaJrASCA}. 
If the \gammaS -ray emission is entirely due to the IC process on CMB
photons, and assuming that the synchrotron and IC emissions are
produced by the same electrons and the emission regions have roughly
the same size ($\xi \approx 1$) then, according to
$w_\gamma/w_{\rm{X}} \simeq 0.1 (B/10\mu \mathrm{G})^{-2} \xi$
\citep{Aharonian97}, the magnetic field in the \gammaS -ray
production region cannot significantly exceed the interstellar value
of several $\mu \mathrm{G}$. If one assumes a larger magnetic field
in the remnant the IC scenario would therefore become less
favourable. On the other hand, the TeV flux can be easily explained in
terms of interactions of accelerated protons with the ambient gas. The
total energy in accelerated protons in the range $10-100$\,TeV
required to provide the observed TeV flux is estimated to be \Wprot,
where $t_{pp
\rightarrow \pi^0} \approx 4.5 \times 10^{15} (n/1\,\mathrm{cm}^{-3})^{-1} \mathrm{s}$ is the
characteristic cooling time of protons through the $\pi^0$ production
channel, and \LGamma\ is the luminosity of the source in \gammaS -rays
between 1 and 10\,TeV. Assuming that the power-law proton spectrum
with spectral index $\alpha \approx \Gamma$ continues down to $E \sim
1 \,\mathrm{GeV}$, the total energy in protons is estimated to be
\Wtot. Thus, for distances to the SNR in the order of $d \approx 200$\,pc 
the conversion of several percent of the assumed mechanical explosion
energy of $10^{51}$\,erg to the acceleration of protons up to $\geq
100$\,TeV would be sufficient to explain the observed TeV
\gammaS -ray flux by nucleonic interactions in a medium of density
comparable to the average density of the interstellar medium, $n \sim
1\,\rm cm^{-3}$. For larger distances a correspondingly higher
fraction of the explosion energy would have to be converted into the
acceleration of protons.

More data will be taken with \hess\ in order to study the morphology of
the remnant in detail, to compare the results with the CANGAROO
measurement and to distinguish between electronic and hadronic
acceleration scenarios.

\begin{acknowledgements}
The support of the Namibian authorities and of the University of Namibia
in facilitating the construction and operation of H.E.S.S. is gratefully
acknowledged, as is the support by the German Ministry for Education and
Research (BMBF), the Max Planck Society, the French Ministry for Research,
the CNRS-IN2P3 and the Astroparticle Interdisciplinary Programme of the
CNRS, the U.K. Particle Physics and Astronomy Research Council (PPARC),
the IPNP of the Charles University, the South African Department of
Science and Technology and National Research Foundation, and by the
University of Namibia. We appreciate the excellent work of the technical
support staff in Berlin, Durham, Hamburg, Heidelberg, Palaiseau, Paris,
Saclay, and in Namibia in the construction and operation of the
equipment.
\end{acknowledgements}

\bibliographystyle{aa}
\bibliography{VelaJr}

\begin{thebibliography}{29}
\expandafter\ifx\csname natexlab\endcsname\relax\def\natexlab#1{#1}\fi

\bibitem[{{Aharonian} {et~al.}(1997){Aharonian}, {Atoyan}, \&
  {Kifune}}]{Aharonian97}
{Aharonian}, F.~A., {Atoyan}, A.~M., \& {Kifune}, T. 1997, \mnras, 291, 162

\bibitem[{{Aharonian} {et~al.}(2004)}]{Nature}
{Aharonian, F.~A. et al. (H.E.S.S.\ collaboration)}. 2004, \nat, 432, 75

\bibitem[{{Aharonian} {et~al.}(2005)}]{PKS2155}
{Aharonian, F. et al. (H.E.S.S.\ collaboration)}. 2005, \aap, 430, 865

\bibitem[{{Aschenbach}(1998)}]{Aschenbach}
{Aschenbach}, B. 1998, \nat, 396, 141

\bibitem[{{Aschenbach} {et~al.}(1999){Aschenbach}, {Iyudin}, \& {Sch{\"
  o}nfelder}}]{Progenitor}
{Aschenbach}, B., {Iyudin}, A.~F., \& {Sch{\" o}nfelder}, V. 1999, \aap, 350,
  997

\bibitem[{{Benbow}(2004)}]{Performance}
{Benbow, W., for the H.E.S.S.\ collaboration}. 2004, in {Proceedings of the
  Gamma 2004 Symposium on High-Energy Gamma-Ray Astronomy}, Vol. 745 (AIP
  Conference Proceedings), 611

\bibitem[{{Bernl{\" o}hr} {et~al.}(2003){Bernl{\" o}hr}, {Carrol}, {Cornils},
  {Elfahem}, {Espigat}, {Gillessen}, {Heinzelmann}, {Hermann}, {Hofmann},
  {Horns}, {Jung}, {Kankanyan}, {Katona}, {Khelifi}, {Krawczynski}, {Panter},
  {Punch}, {Rayner}, {Rowell}, {Tluczykont}, \& {van Staa}}]{StatusOptics}
{Bernl{\" o}hr}, K., {Carrol}, O., {Cornils}, R., {et~al.} 2003, Astroparticle
  Physics, 20, 111

\bibitem[{{Cornils} {et~al.}(2003){Cornils}, {Gillessen}, {Jung}, {Hofmann},
  {Beilicke}, {Bernl{\" o}hr}, {Carrol}, {Elfahem}, {Heinzelmann}, {Hermann},
  {Horns}, {Kankanyan}, {Katona}, {Krawczynski}, {Panter}, {Rayner}, {Rowell},
  {Tluczykont}, \& {van Staa}}]{StatusOptics2}
{Cornils}, R., {Gillessen}, S., {Jung}, I., {et~al.} 2003, Astroparticle
  Physics, 20, 129

\bibitem[{{Drury} {et~al.}(1994){Drury}, {Aharonian}, \& {V{\"
  o}lk}}]{DruryAharonianVoelk94}
{Drury}, L.~O., {Aharonian}, F.~A., \& {V{\" o}lk}, H.~J. 1994, \aap, 287, 959

\bibitem[{{Duncan} \& {Green}(2000)}]{VelaJrRadio}
{Duncan}, A.~R. \& {Green}, D.~A. 2000, \aap, 364, 732

\bibitem[{{Enomoto} {et~al.}(2002){Enomoto}, {Tanimori}, {Naito}, {Yoshida},
  {Yanagita}, {Mori}, {Edwards}, {Asahara}, {Bicknell}, {Gunji}, {Hara},
  {Hara}, {Hayashi}, {Itoh}, {Kabuki}, {Kajino}, {Katagiri}, {Kataoka},
  {Kawachi}, {Kifune}, {Kubo}, {Kushida}, {Maeda}, {Maeshiro}, {Matsubara},
  {Mizumoto}, {Moriya}, {Muraishi}, {Muraki}, {Nakase}, {Nishijima}, {Ohishi},
  {Okumura}, {Patterson}, {Sakurazawa}, {Suzuki}, {Swaby}, {Takano}, {Takano},
  {Tokanai}, {Tsuchiya}, {Tsunoo}, {Uruma}, {Watanabe}, \&
  {Yoshikoshi}}]{Enomoto}
{Enomoto}, R., {Tanimori}, T., {Naito}, T., {et~al.} 2002, \nat, 416, 823

\bibitem[{{Funk} {et~al.}(2004){Funk}, {Hermann}, {Hinton}, {Berge}, {Bernl{\"
  o}hr}, {Hofmann}, {Nayman}, {Toussenel}, \& {Vincent}}]{CentralTrigger}
{Funk}, S., {Hermann}, G., {Hinton}, J., {et~al.} 2004, Astroparticle Physics,
  22, 285

\bibitem[{{Iyudin} {et~al.}(2005){Iyudin}, {Aschenbach}, {Becker}, {Dennerl},
  \& {Haberl}}]{XMM}
{Iyudin}, A.~F., {Aschenbach}, B., {Becker}, W., {Dennerl}, K., \& {Haberl}, F.
  2005, \aap, 429, 225

\bibitem[{{Iyudin} {et~al.}(1998){Iyudin}, {Sch{\" o}nfelder}, {Bennett},
  {Bloemen}, {Diehl}, {Hermsen}, {Lichti}, {van der Meulen}, {Ryan}, \&
  {Winkler}}]{Iyudin}
{Iyudin}, A.~F., {Sch{\" o}nfelder}, V., {Bennett}, K., {et~al.} 1998, \nat,
  396, 142

\bibitem[{{Kargaltsev} {et~al.}(2002){Kargaltsev}, {Pavlov}, {Sanwal}, \&
  {Garmire}}]{CHANDRACentral}
{Kargaltsev}, O., {Pavlov}, G.~G., {Sanwal}, D., \& {Garmire}, G.~P. 2002,
  \apj, 580, 1060

\bibitem[{{Katagiri} {et~al.}(2005){Katagiri}, {Enomoto}, {Ksenofontov},
  {Mori}, {Adachi}, {Asahara}, {Bicknell}, {Clay}, {Doi}, {Edwards}, {Gunji},
  {Hara}, {Hara}, {Hattori}, {Hayashi}, {Itoh}, {Kabuki}, {Kajino}, {Kawachi},
  {Kifune}, {Kiuchi}, {Kubo}, {Kurihara}, {Kurosaka}, {Kushida}, {Matsubara},
  {Miyashita}, {Mizumoto}, {Muraishi}, {Muraki}, {Naito}, {Nakamori}, {Nakase},
  {Nishida}, {Nishijima}, {Ohishi}, {Okumura}, {Patterson}, {Protheroe},
  {Sakamoto}, {Sakamoto}, {Swaby}, {Tanimori}, {Tanimura}, {Thornton},
  {Tsuchiya}, {Watanabe}, {Yamaoka}, {Yanagita}, {Yoshida}, \&
  {Yoshikoshi}}]{Cangaroo2}
{Katagiri}, H., {Enomoto}, R., {Ksenofontov}, L.~T., {et~al.} 2005, \apjl, 619,
  L163

\bibitem[{{Koyama} {et~al.}(1997){Koyama}, {Kinugasa}, {Matsuzaki},
  {Nishiuchi}, {Sugizaki}, {Torii}, {Yamauchi}, \& {Aschenbach}}]{RXJ1713_ASCA}
{Koyama}, K., {Kinugasa}, K., {Matsuzaki}, K., {et~al.} 1997, \pasj, 49, L7

\bibitem[{{Koyama} {et~al.}(1995){Koyama}, {Petre}, {Gotthelf}, {Hwang},
  {Matsuura}, {Ozaki}, \& {Holt}}]{SN1006_ASCA}
{Koyama}, K., {Petre}, R., {Gotthelf}, E., {et~al.} 1995, \nat, 378, 255

\bibitem[{{Li} \& {Ma}(1983)}]{LiMa}
{Li}, T.-P. \& {Ma}, Y.-Q. 1983, \apj, 272, 317

\bibitem[{{Masterson} {et~al.}(2004)}]{HESSCrab}
{Masterson, C., et al. for the H.E.S.S.\ collaboration}. 2004, in {Proceedings
  of the Gamma 2004 Symposium on High-Energy Gamma-Ray Astronomy}, Vol. 745
  (AIP Conference Proceedings), 617

\bibitem[{{Mereghetti}(2001)}]{VelaJrBeppoSAX}
{Mereghetti}, S. 2001, \apjl, 548, L213

\bibitem[{{Muraishi} {et~al.}(2000){Muraishi}, {Tanimori}, {Yanagita},
  {Yoshida}, {Moriya}, {Kifune}, {Dazeley}, {Edwards}, {Gunji}, {Hara}, {Hara},
  {Kawachi}, {Kubo}, {Matsubara}, {Mizumoto}, {Mori}, {Muraki}, {Naito},
  {Nishijima}, {Patterson}, {Rowell}, {Sako}, {Sakurazawa}, {Susukita},
  {Tamura}, \& {Yoshikoshi}}]{CangarooRXJ}
{Muraishi}, H., {Tanimori}, T., {Yanagita}, S., {et~al.} 2000, \aap, 354, L57

\bibitem[{{Pavlov} {et~al.}(2001){Pavlov}, {Sanwal}, {K{\i}z{\i}ltan}, \&
  {Garmire}}]{VelaJr2}
{Pavlov}, G.~G., {Sanwal}, D., {K{\i}z{\i}ltan}, B., \& {Garmire}, G.~P. 2001,
  \apjl, 559, L131

\bibitem[{{Pellizzoni} {et~al.}(2002){Pellizzoni}, {Mereghetti}, \& {De
  Luca}}]{HaCenter}
{Pellizzoni}, A., {Mereghetti}, S., \& {De Luca}, A. 2002, \aap, 393, L65

\bibitem[{{Slane} {et~al.}(2001){Slane}, {Hughes}, {Edgar}, {Plucinsky},
  {Miyata}, {Tsunemi}, \& {Aschenbach}}]{VelaJrASCA}
{Slane}, P., {Hughes}, J.~P., {Edgar}, R.~J., {et~al.} 2001, \apj, 548, 814

\bibitem[{{Tsunemi} {et~al.}(2000){Tsunemi}, {Miyata}, {Aschenbach}, {Hiraga},
  \& {Akutsu}}]{Tsunemi}
{Tsunemi}, H., {Miyata}, E., {Aschenbach}, B., {Hiraga}, J., \& {Akutsu}, D.
  2000, \pasj, 52, 887

\bibitem[{{V{\" o}lk} {et~al.}(2005){V{\" o}lk}, {Berezhko}, \&
  {Ksenofontov}}]{Voelk}
{V{\" o}lk}, H.~J., {Berezhko}, E.~G., \& {Ksenofontov}, L.~T. 2005, \aap, 433,
  229

\bibitem[{{Vincent} {et~al.}(2003){Vincent}, {Denance}, \& {Huppert, J.-F. et
  al.}}]{StatusCamera}
{Vincent}, P., {Denance}, J.-P., \& {Huppert, J.-F. et al.} 2003, in Proc. 28th
  ICRC, Tsukuba (Univ. Academy Press, Tokyo), 2887

\bibitem[{{Voges} {et~al.}(1999){Voges}, {Aschenbach}, {Boller}, {Br{\"
  a}uninger}, {Briel}, {Burkert}, {Dennerl}, {Englhauser}, {Gruber}, {Haberl},
  {Hartner}, {Hasinger}, {K{\" u}rster}, {Pfeffermann}, {Pietsch}, {Predehl},
  {Rosso}, {Schmitt}, {Tr{\" u}mper}, \& {Zimmermann}}]{ROSAT}
{Voges}, W., {Aschenbach}, B., {Boller}, T., {et~al.} 1999, \aap, 349, 389

\end{thebibliography}

\end{document}